\documentclass{amsart}

\usepackage{graphicx}
\usepackage{epstopdf}
\usepackage{tikz,xcolor}
\usepackage{multirow}
\usepackage{float}

\usepackage{amsthm,amssymb}

\newtheorem{thm}{Theorem}[section]

\theoremstyle{definition}
\newtheorem*{defn}{Definition}

\newcommand{\ignore}[1]{}

\def\hpic #1 #2 {\mbox{$\begin{array}[c]{l} \epsfig{file=#1,height=#2} \end{array}$}}
\def\vpic #1 #2 {\mbox{$\begin{array}[c]{l} \epsfig{file=#1,width=#2} \end{array}$}}

\newcommand{\znote}[1]{}
\newcommand{\fnote}[1]{}

\begin{document}

\title{Fair division and Redistricting}

\author{Zeph Landau}
\address{Department of Engineering and Computer Science \\ University of California at Berkeley \\Berkeley CA 94720 }
\email{zeph.landau@gmail.com}

\author{Francis Edward Su}
\address{Department of Mathematics\\ Harvey Mudd College\\ Claremont, CA  91711}
\email{su@math.hmc.edu}


\thanks{Research partially supported by NSF Grants DMS-0701308 and DMS-1002938 (Su).}


\dedicatory{\ \ \ }
\maketitle

\section{Introduction}

Redistricting is the political practice of dividing states into electoral districts of equal population.  It is mandated to occur every ten years, after the census, to ensure equal representation in the legislative body. Where the boundaries 
are drawn can dramatically alter the number of districts a given political 
party can win. As a result, a political party which has control over the 
legislature, can (and does) manipulate the boundaries to win a larger number of districts, thus affecting the balance of power in the U.S. House of 
Representatives.  This kind of boundary manipulation occurs even with certain legal and legislative constraints that restrict some aspect of how districts can be drawn and mandate that, where appropriate, districts should be created to have a majority of voters consisting of a racial minority. (See \cite{redlaw2000} for a detailed summary of these constraints.)

The ability of one political party to gain political advantage by carefully choosing the boundaries during redistricting has been recognized as a serious problem with the redistricting process in the United States; we shall refer to this as the problem of {\it partisan unfairness}.  Attempts to fix and/or mitigate the problem of partisan unfairness (beyond the legal restrictions) have taken one of two approaches: trying to constrain the process to reduce the amount of political gain achievable, and trying to remove politics from the redistricting process.  Examples of the first approach include attempting to limit the power of the drawing party by more strictly prescribing the allowable shapes of districts, and banning the use of registration and voting data within the redistricting process.  Examples of the second approach include assigning the task of redistricting to bipartisan or non-partisan panels\footnote{Eleven states redistrict using bipartisan or independent  commissions.},  and using computer programs to generate redistricting maps that optimize certain carefully chosen criteria.

Landau-Reid-Yershov \cite{landau} took a different approach to provide a novel solution to the problem of partisan unfairness:  rather than trying to fix the problem by restricting the shape of the possible maps or by assigning the power to draw the map to nonbiased entities, their solution ensures fairness by balancing competing interests against each other.   
This kind of solution is an example of what are known as ``fair division'' solutions--- 
such solutions account explicitly for the preferences of all parties, are determined by a procedure in which all parties are actively involved, and are accompanied by rigorous guarantees of a specified notion of fairness.  

The goal of this article is to provide an exposition of this redistricting method in the context of a detailed sample ``map'', and make a stronger connection to the ideas of fair division than is provided in \cite{landau}.  In particular, we propose a specific notion of fairness that was used but not made explicit in \cite{landau}.  This notion of fairness can be used in concert with (not substitute for) other necessary or desired criteria for a good redistricting solution.  We clarify how fair division ideas can play an important role in a realistic redistricting solution by introducing an interactive step that involves multilateral evaluation, procedural fairness, and fairness guarantees.
And by making the bridge between fair division ideas and redistricting solutions more explicit, we hope to encourage the flow of ideas between the two areas.

We begin with an introduction to the ideas of fair division in Section \ref{s:2}.  We describe the problem of partisan unfairness in more detail in Section \ref{s:3}.  Despite the ability to easily recognize the ``unfairness'' inherent in the redistricting process, it has been hard to give a reasonable definition of what would be fair.  In Section \ref{s:4} we give an explicit definition of fairness---\emph{the geometric target}---that incorporates geometric considerations such as how constituent voters are distributed and how districts are shaped.
Section \ref{s:6} examines the protocol of \cite{landau} by analyzing its behavior in a specific example that demonstrates most aspects of the solution.   Within this example, the fairness of the protocol is discussed in detail.

We stress that the ideas discussed here are well suited to be combined with other necessary or desired criteria for a good redistricting solution.   Our definition of fairness involving geometric targets in Section \ref{s:4} 
can incorporate independently desired requirements for district shape or competitiveness.  The protocol of \cite{landau} can be easily adjusted to approximate fairness under these additional requirements.  If a solution involving an independent commission is desired, these definitions of fairness can be used as a target for the commission.  Similarly if a computer assisted solution that consists of optimizing some function is sought, this measure of fairness can be used as a component of the function to be optimized.  Separately, the simple fair division ranking protocol given in Section \ref{s:5} can be used to incorporate some degree of legislative preference within any proposed redistricting solution that otherwise does not include legislative influence.

\section{Fair Division}
\label{section-fair} \label{s:2}

The problem of {\em fair division}, as Steinhaus put it \cite{steinhaus} is essentially a question of how to divide some object {\em fairly}.  Usually, this object is affectionately referred to as {\em cake} \cite{robertson-webb}, but in general it could be desirable or undesirable (e.g., the division of {\em chores \cite{peterson-su}}) or a mixture of desirable and undesirable goods \cite{su}.  The cake may be infinitely divisible (as we usually regard real cake) or only divisible into discrete pieces (such as a pieces of an estate).  Applications of fair division ideas include methods for resolving international disputes and divorce settlements \cite{brams-taylor-book}.

There are several notions of {\em fairness} that one might consider, but an important aspect of fair division problems is that this fairness notion is evaluated by the parties involved in the negotation, rather than an outside arbiter.  Thus, the outcome of a fair division procedure will give each party what it considers to be a fair share, according to its own evaluation.

The simplest example of a fair division procedure is the familiar ``I-cut-you-choose'' method for dividing a cake among two people.  One might consider a fair piece in which each party does not envy the other; we call such a solution an {\em envy-free} solution.  Again, note that envy is measured by each party according to its own evaluation.  If one person cuts the cake (into two pieces that she is indifferent between) and the other person is allowed to choose first (picking the piece that he most desires), then both people will end up with a piece for which they experience no envy.  This simplest of all fair division procedures already highlights some interesting features common to all fair division procedures:

\begin{enumerate}
\item
{\em Multilateral evaluation.}  Fairness is evaluated according to each party's own preferences.  Therefore, parties don't have to agree on what is valuable; each will obtain a share they would consider fair in their own estimation (and they do not need to know the other party's preferences).

\item
{\em Procedural fairness.}   There is a process by which preferences are elicited, all parties are involved in the process, and they understand the criteria by which fairness is measured.  Because of this, parties are more likely to feel that the process is fair, more so than a decision imposed by an outside arbiter (see e.g., \cite{spector}).  The procedure guides parties to a mutually acceptable division.

\item
{\em Fairness guarantee.}  
By following the procedure, as long as you tell the truth about your preferences, you will obtain what you feel is fair (even if everyone else lies).
Thus there is an incentive to be truthful (and if you lie about your preferences, it can backfire).
\end{enumerate}

A reader might object to the above particular cake-division solution, because the cutter only gets what he perceives to be 50 percent of the cake in his valuation, while the other person might end up with more.  The solution is envy-free (neither person envies the other person's share) but it is not 
{\em equitable}, meaning the perceived share of cake each player gets (in his own valuation) is different. This is not a deficit of the procedure (which only guaranteed envy-freeness and not equitability) as much as it is a deficit of the procedure chosen.  An active area of research in mathematics \cite{robertson-webb}, economics \cite{moulin}, and political science \cite{brams-taylor-book} is the development of fair division procedures in various settings and with various fairness criteria.

A more interesting fair division solution that has found application by practitioners is the Adjusted Winner procedure of Brams and Taylor \cite{brams-taylor-win}.  It is procedure for dividing a set of goods between two parties in such a way that the division is: envy-free, equitable, and {\em efficient} (or {\em Pareto-optimal}).  The last criterion means that there is no division that dominates the given one, i.e., there is no other division that is just as good for both parties and strictly better for one party.  Thus the Adjusted Winner solution gives each party a share for which they do not wish to trade shares, and in which they feel they got just as good a portion as the other party feels it got, and there is no other solution that dominates.  At most one of the goods may have to be divided in the procedure (though one cannot predict beforehand which procedure it is).  We note that if there are more than two parties, it may not always be possible to satisfy these three properties in cake division, as discussed in \cite{brams-jones-klamler}.

The Adjusted Winner procedure has found application in divorce settlements \cite{brams-taylor-win} because of its fairness guarantees as well as its ease of use.  In the procedure, both parties are given 100 points to divide by assigning over the objects.  This is the part of the procedure where preferences are elicited.  Then objects are initially given to the party that valued them most; such a division is efficient, but it may not be envy-free or equitable.  Call the party who ends up with the largest fractional share (in its own evaluation) the {\em winner}, and the other, the {\em loser}.  In the next phase of the procedure, the assignment of goods is ``adjusted'' by transferring goods from the winner to the other party in a particular order until both fractional shares are equalized.  

The Adjusted Winner procedure has the 3 features described above for a fair division procedure.
It has a {\em fairness guarantee}: what results is an outcome that is provably envy-free and efficient, in addition to being equitable. 
It relies on {\em mutilateral evaluation}: the preferences of both parties are taken into account, and the resulting division achieves the fairness guarantee for both parties using their own estimation.  And it is {\em procedurally fair}: parties using the Adjusted Winner procedure can understand and verify the fairness guarantees for a particular solution (without having to understand the proofs); because they participated in the procedure by stating their preferences, they are more likely to feel that the outcome is fair.

As we shall see, these three fair division ideas can offer some helpful ideas to current thinking about redistricting which can be combined with other desired ideas for a good redistricting solution.  They underlie the redistricting procedure of \cite{landau} that we will now explain.  First, we will explain the problem of partisan unfairness that \cite{landau} attempts to address.

\section{Redistricting: the problem of partisan unfairness} \label{s:3}

In most of the 50 states in the U.S., the districting protocol is to have one party draw all the boundaries.   If the drawing party's goal is to win as many districts as it can, the strategy is clear: draw boundaries so that each district either 
a) has a small majority of its voters, or 
b) has a large majority of the other party's voters.  
In other words, for any district, the drawing party should strive to either win it by a small margin or lose it by a large margin (See \cite{app} for a detailed discussion).  

In general, with such a strategy,
a drawing party with $X \% $ of support of the voters 
can win just under $\min (2X \%, 100\%)$ of the districts if there are no geometric constraints
(e.g., requiring districts be compact or contiguous, etc.).
In reality, geometric constraints usually mean that this ideal outcome cannot be achieved; 
however, in most cases, the drawing party can still win a significantly larger percentage than $X \%$ of the districts, even with only partial knowledge of voting trends.

This is not just a theoretical issue, as has often been demonstrated when the party in control changes.  
We cite two examples:
\begin{itemize}
\item
When Republicans took control of the Texas legislature in 2002, they redrew state districts mid-decade, and the Texas delegation changed from 15 Republicans and 17 Democrats
to 22 Republicans and 10 Democrats. \cite{electstat}
\item 
In Michigan, the 2000 election produced 7 Republican representatives and 9 Democratic representatives.  After the census, a new district map was drawn resulting in 9 Republican representatives and 6 Democratic representatives in the 2002 election (Michigan lost 1 seat due to the census).\cite{electstat}
\end{itemize}

This ability of one party to draw districts in such a way as to gain political advantage is viewed 
as one of the major problems with redistricting in the United States; 
we shall refer to this as the problem of \emph{partisan unfairness}. 
The districting protocol proposed in \cite{landau} avoids 
this inherent unfairness by ensuring that either party 
can win a percentage of districts that is very close to their fair share.

\section{What is a party's fair share?} \label{s:fairshare} \label{s:4}

Defining a ``fair'' share is not as straightforward as it may seem.
A reasonable first attempt would be to say that the percentage of districts won by a party 
should be close to the percentage of constituent voters in the party.  
However, geometric constraints can make this impossible.  For instance, 
if one party enjoys a statewide 60\% - 40\% advantage in the electorate 
and if the parties are mixed homogeneously throughout the state, then
any reasonable district would have a similar 60\% - 40\% advantage for the same party.
In such a case, if we believe that districts should contain people 
who live in contiguous and relatively compact regions, then all districts would go
to the majority party and we would consider this outcome fair even 
though it differed dramatically from the percentage of constituent voters.
This example demonstrates the need to incorporate geometric constraints into any reasonable notion of fair shares.

It may come as a surprise that such a fairness notion can be defined, despite the complexity of possible geometric constraints.
In this section, we define a fairness notion that incorporates geometric considerations such as 
how constituent voters are distributed and how districts are shaped.  Before doing so, we make some preliminary definitions.

Any division of a state into districts that satisfies all desirable constraints (including legal constraints that district maps are subject to) will be called a \emph{viable division}.
The {\it voting outcome} $V_{out}$ is a description of how every voter actually votes in a given election. Any party with a role in redistricting attempts to predict aspects of $V_{out}$ to guide their choices of district boundaries.  A {\it voting model} $V$ shall be such a prediction of how every voter will vote.

A party's {\it rating system} $R$ is an assignment of a number to any proposed 
viable division of a state that gives a measure of how desirable that viable division is.
A simple example of a rating system is to rate a viable division by the number of districts the party thinks it can win; we will denote this rating system by $R_{win}$.  Implicit in calculating 
$R_{win}$ is a voting model $V$ that the party is using to predict which districts it will win.

In reality, a party's interests may be much more complicated then 
just the number of districts that it wins.   A more general rating system is one where a party could rate the desirability of each district in a division (assigning it a number), then sum these numbers over all the districts to give a rating for the division.  As in  \cite{landau}, we shall refer to such a rating sytem as {\it an additive rating system},  and denote any particular instance of it as $R_{sum}$.  The rating system $R_{win}$ is a special case of $R_{sum}$ in which
a party assigns a $1$ to districts it expects to win and a $0$ to those it expects to lose. 

The more general rating system $R_{sum}$ allows a party to take other considerations into account.  
Politically, these can be important, as the following examples demonstrate:
\begin{itemize}
\item
perhaps some district has an incumbent who is on an important congressional committee, so winning that district is more valuable to the party (hence rated higher than other winnable districts),
\item 
perhaps some district has an important landmark (a stadium or a construction project)
worth more to a party than some other district,
\item perhaps some district encompasses the supporters of two 
incumbents from the opposition party, so that even though the district will be lost, 
the elimination of one strong incumbent from the other party is valuable.
\end{itemize}

Equipped with the notions of viable division, voting model, and rating system, we can now define our fairness notion:

\begin{defn} [Geometric Target] Given a voting model $V$ and a rating system $R$, a party's {\it geometric target with respect to $V$ and $R$} is the 
average of its highest and lowest ratings among the set of viable divisions.  A party's geometric target with respect to $V_{out}$ and $R_{win}$ 
will be called the {\it absolute geometric target}.
\end{defn}

The absolute geometric target is a definition of fairness that takes into account geometric constraints.   There are several compelling reasons why this is a good definition.  First, the definition seems conceptually fair as it lies exactly between the best and worst outcome (in terms of number of districts won) for each party.  Second, when there are no geometric constraints, the absolute geometric target coincides with the percentage of constituent voters---as already mentioned, if the minority party has $X\%$ of the vote, its best outcome is  to win about $2X \%$ of the districts, while its worst outcome is to lose all the districts $0 \%$ and so the absolute geometric target in this case would be approximately $\frac{2X +0}{2}= X \%$ of the districts.
Third, because this definition uses only viable district maps, it incorporates geometric constraints by restricting attention to realizable outcomes.  For instance, in the example at the beginning of this section (with the homogeneous 60\% - 40\% electorate split), both the best and worst rating for the minority party would be to win $0$ districts which is the only possible outcome (and thus fair) in that case.  We remark that the absolute geometric target could be used as a target for fairness for independent commissions.

The protocol for districting proposed in \cite{landau} (and described subsequently), allows each party the opportunity to achieve an outcome that is 
close to their own absolute geometric target.
Moreover, it allows each party the opportunity to achieve an outcome that is 
close to a geometric target with respect to any voting model $V$ and any additive 
rating system $R$.


Note that the geometric target with respect to a voting model $V$ and a rating system $R$ is a notion that captures the fair division principle of {\em multilateral evaluation}: 
that party preferences should be taken into account.  
Each party has its own geometric target, based on its own voting model $V$ and rating system $R$(derived from its preferences).  This is to be distinguished from any absolute notions of fairness that might be imposed by an external arbiter (including the absolute geometric target).  

We shall soon see that districting protocol of \cite{landau} will, in addition, possess the other features of a fair division procedure--- {\em procedural fairness}, and a {\em fairness guarantee} (see Section \ref{section-fair}).

\section{The ranking protocol} \label{ranking} \label{s:5}
Before presenting the redistricting protocol of \cite{landau},  we present a simple but useful protocol, which we call the {\it ranking protocol}, for how two parties can decide on one of a bunch of outcomes.    For our purposes, the setting will be that of two political parties, $A$ and $B$, where the choice of outcomes are different proposed divisions of a state.  The protocol is then simple: both parties rank the proposed divisions from best to worst from their perspective.  Each proposed division then has two rankings. 
Select the proposed division whose worse ranking is best (this reflects the Rawls' maximin criterion, as discussed in \cite{rawls}). If there are 2 such proposals (there can be at most 2), randomly choose one.

Notice that if there are $n$ proposed divisions, the division chosen is guaranteed to be in the top $\frac{n}{2} +1$ of both lists.  Said another way, the ranking protocol provides an outcome that, for either party, is no worse than one less than their median outcome among the choices (and can be much better if the two parties desires are not diametrically opposed).  As we shall see in the next section, the ranking protocol is used as an augmentation step for the core redistricting protocol of \cite{landau}.  

We point out more generally that the ranking protocol could be used with any proposed method for generating divisions of a state, be it divisions created by independent panels, computers, or based on any kind of optimization scheme.

\section{The fair division redistricting protocol} \label{s:6}

We now describe the core protocol for redistricting that was presented in \cite{landau}.  It will involve three parties: two parties with vested interest in the division (e.g. the democratic and republican parties, or the majority and minority party in a state) called parties $A$ and $B$, and an independent agent which we'll refer to as $I$.

The formal precise description of the protocol can be found in \cite{landau}; here we will describe the protocol  while working through an example that will illustrate many of the aspects of both the protocol and the resulting solution.

\subsection{The State}

\begin{figure}[H] 
\includegraphics[height=2in]{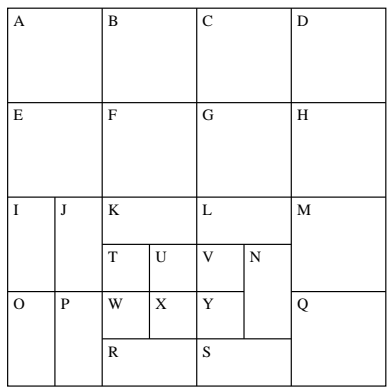}
\caption{Sample map of districts.}
\label{map-letters}
\end{figure}

Consider the map in Figure \ref{map-letters}, which represents a state consisting of 25 parcels, which we are thinking of as indivisible units (here, they are rectangles or squares).  
Each parcel contains the same number of people; thus the smaller the parcel, the denser the population.
Loosely, we can think of this state as having a city located at the small rectangles (T,U,V,W,X,Y), with suburban areas surrounding the city, and the remaining areas rural.

Suppose our goal is to divide this state into 5 districts, each containing exactly 5 parcels. 

\begin{figure}[H]   
\includegraphics[height=2in]{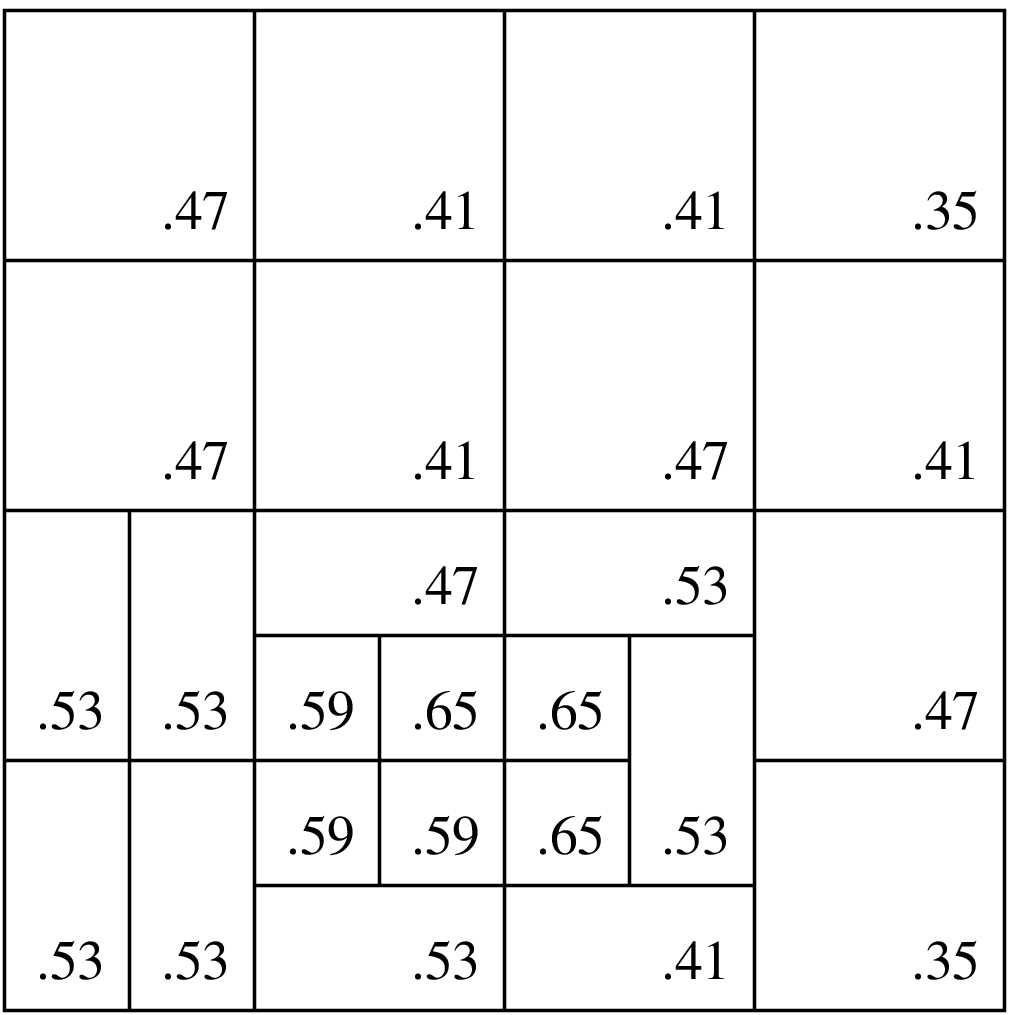}
\caption{Vote totals by district.}
\label{map}
\end{figure}

Suppose that the voting outcome ($V_{out}$) of the ensuing election is given by Figure \ref{map}, which shows in each parcel the percentage of votes that party $A$ receives.

  The ability to use the redistricting process to gain political advantage relies on the ability to predict some features of $V_{out}$.
In general, of course, $V_{out}$ is not known precisely at the time districts are being re-drawn.
However, 
the combination of data from previous elections and opinion polling increasingly gives a 
more and more accurate model of how votes will be distributed. 
For the purposes of this example we will assume that the working voting model $V$ of both parties coincides with $V_{out}$ in Figure \ref{map}.
In this example, party $A$ has a slim statewide majority, receiving 50.12\% of the total vote.


For this particular example, we will assume that the only thing the two parties care about is maximizing the number of districts they can win; thus their preferences are diametrically opposed with each having rating system $R_{win}$.  We emphasize that this is an assumption we make for this example but that the protocol is designed to work under much more general preferences---the additive ratings system  $R_{sum}$, discussed earlier.

\begin{figure}[H] 
\includegraphics[height=2in]{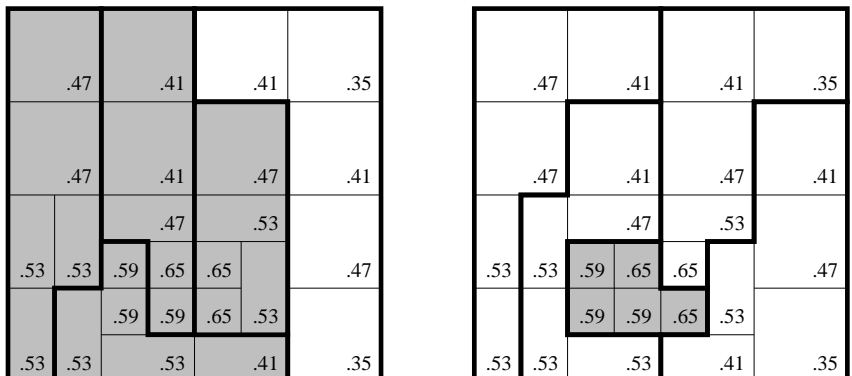}
\caption{The left diagram shows a division in which $A$ can win 4 districts.  
The right diagram shows a division in which $B$ can win 4 districts.  Districts that $A$ wins are shaded.}
\label{map-4of5}
\end{figure}

Notice that even though parties $A$ and $B$ each have approximately half the voters over the state, 
if either party is given complete control of the district-drawing process, they can draw districts so that they are the majority in 4 of the 5 districts.  See Figure \ref{map-4of5}.
In this example, the absolute geometric target for either party is $\frac{1 +4}{2}= 2.5$ districts.

\subsection{The Protocol} \label{s:protocol}
There are three core steps to the redistricting protocol along with an augmenting fourth step.   After outlining them in general, we will work through each step in the above example.
\begin{itemize}
\item {\bf Split Sequence Generation.}  This step is performed by the independent agent $I$.  The agent generates a sequence of so-called  $k$-splits: a $k$-split is a division of the state into two pieces  (piece 1 and piece 2) such that the population within piece 1 totals the number of people in $k$ districts.  The independent agent $I$ generates a {\it split sequence}: a 1-split, a 2-split, a 3-split, etc. with each split building on the previous so that piece 1 of the  $j$-split contains piece 1 of the $(j-1)$-split for all $j$.

\item{\bf Preference.}  For each of the $k$-splits, the two parties are each asked which of the following options they would prefer:

\begin{enumerate}
\item to have party $A$ divide piece 1 of the split into $k$ districts and have party $B$ divide piece 2 into $n-k$ districts (where $n$ is the total number of districts).
\label{optiona}
\item  to have party $B$ divide piece 1 of the split into $k$ districts and have party $A$ divide piece 2 into $n-k$ districts. \label{optionb}
\end{enumerate}

Each party has the option of saying that they are indifferent to the two choices\footnote{This additional option is a modification of the original protocol in \cite{landau}.}.

\item{\bf Resolution.}  If there exists an $i$-split such that parties $A$ and $B$ both prefer the same option in the preference step above  then create a map using that option.   If there exists an $i$-split such that one party is indifferent, then create a map using the option selected by the party that was not indifferent.  If there exists an $i$-split such that both parties are indifferent, then create a map  by randomly choosing one of the options for that $i$-split.
\smallskip

 If none of the above scenarios occur it means that the parties have opposite preferences for each $i$.  Find the first  $i_0$, $1 \leq i_0 \leq n-2$ for which party $A$ prefers option (\ref{optionb}) for $i=i_0$ and  switches preferences to option (\ref{optiona}) when  $i=i_0 +1$.  (This scenario is guaranteed to occur at least once since party $A$ prefers option (\ref{optionb}) when $i=1$ and prefers option (\ref{optiona}) when $i=n-1$.)  Randomly choose to divide the state from the following four prescriptions:
\begin{enumerate}
\item[i.]  use option (\ref{optiona}) for the $i_0$-split, 
\item[ii.] use option (\ref{optionb}) for $i_0$-split, 
\item[iii.] use option (\ref{optiona}) for $(i_0 +1)$-split,
\item[iv.] use option (\ref{optionb}) for $(i_0 +1)$-split.
\end{enumerate}

\item{\bf Augmentation.}   We perform the above 3 steps for a number of different split sequences to produce a number of divisions of the state.  We then use the ranking protocol of Section \ref{ranking} to choose among these maps, i.e. 
each party  ranks the divisions (from best to worst) according to their own preferences, and the division whose worst ranking (among both parties) is highest is the one that is chosen. If there are two such splits, select one of them at random.

\end{itemize}

\bigskip

\subsection{The protocol in action.}
We now show how the protocol works for the example described in Figure \ref{map}.

\begin{figure}[H]
\includegraphics[height=1.25in]{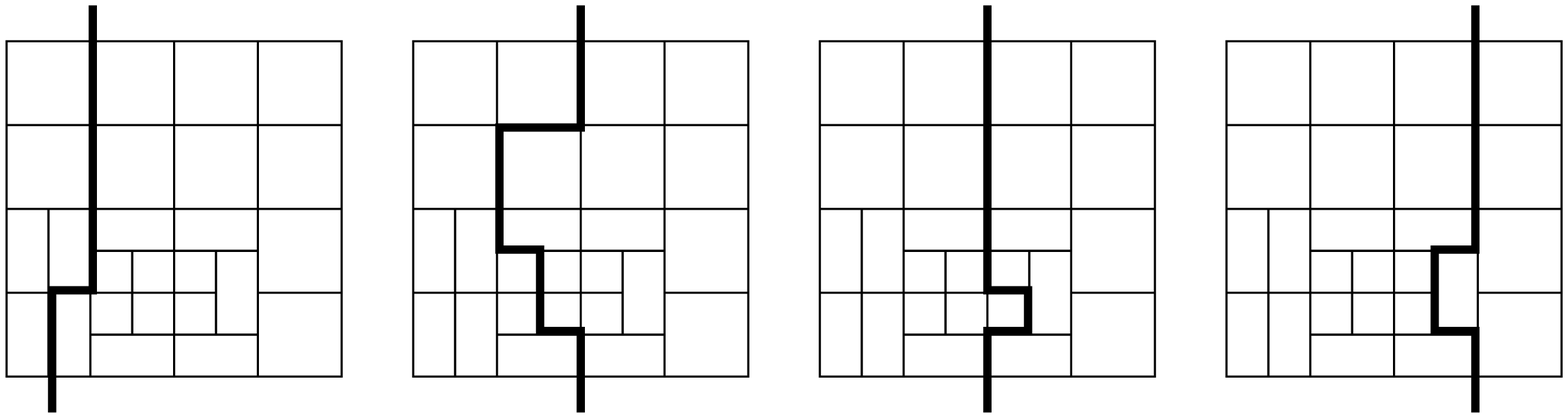}
\caption{Split sequence generation.  From left to right, a $1$-split, $2$-split, $3$-split, $4$-split.}
\label{newsplit}
\end{figure}

{\it Split Sequence Generation step.} Suppose the Split Generation step yields the split sequence in Figure \ref{newsplit}.
In each diagram piece 1 will be the left piece and piece 2 will be the right piece.

\begin{figure}[H]
\includegraphics[height=2in]{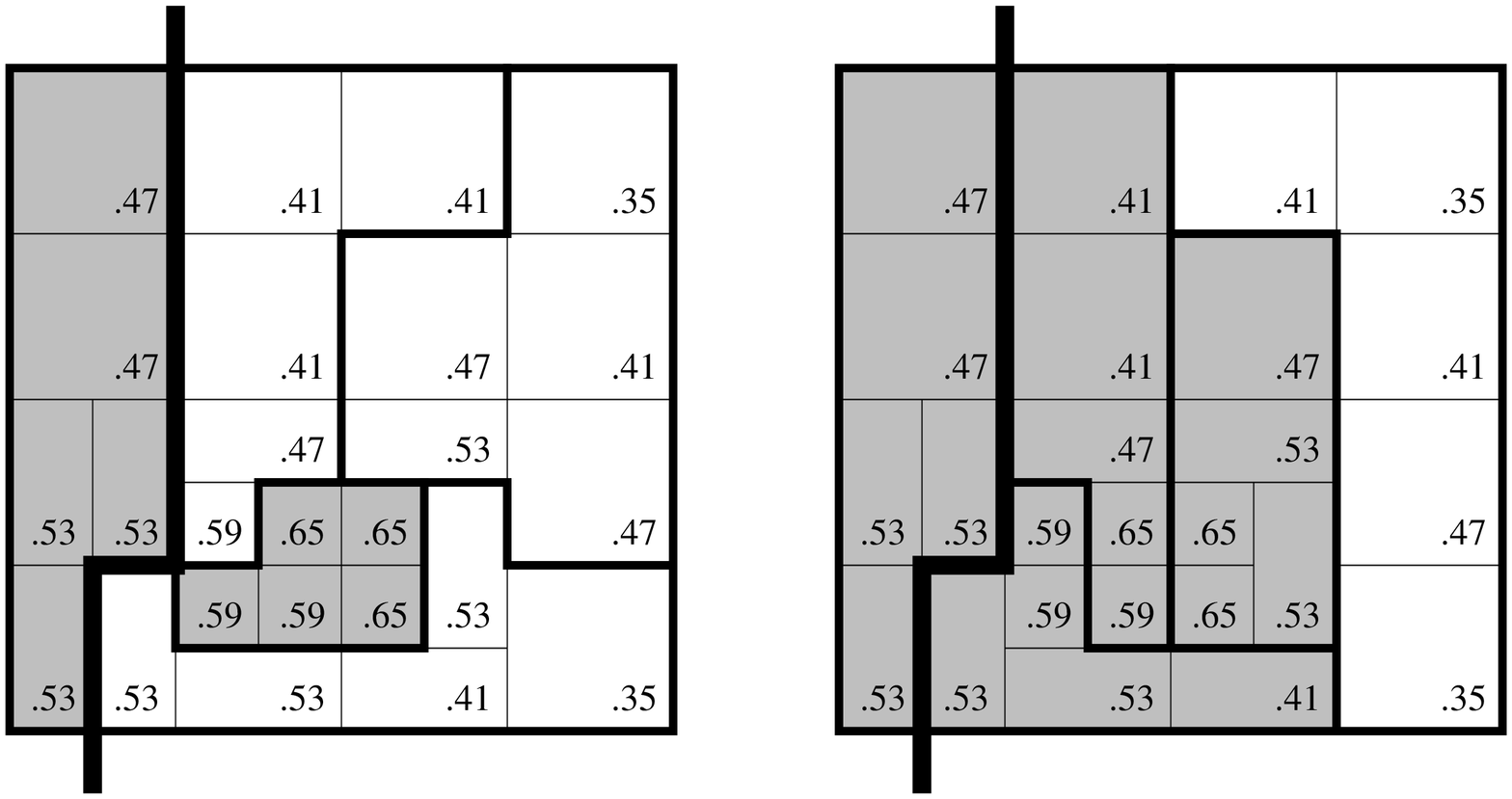}
\caption{Two options for the $1$-split.  The first diagram is option (\ref{optiona}): 
in which $A$ divides the left piece and $B$ divides the right piece.  
The second diagram is option (\ref{optionb}) in which $B$ divides the left piece and $A$ divides the right piece.}
\label{1-split}
\end{figure}

{\it Preference step.}  For the 1-split,  we exhibit in Figure \ref{1-split} a possible way each party can optimize its own interests in each of the two options.  In option (\ref{optiona}), $A$ divides piece 1 and $B$ divides piece 2.  In option (\ref{optionb}), $B$ divides piece 1 and $A$ divides piece 2.

Thus party $A$ would prefer option (\ref{optionb}) (since it would win 4 out of 5 districts) and party $B$ would prefer option (\ref{optiona}) (since it would win 4 out of 5 districts).   This is not surprising since there is no opportunity to gerrymander the left piece.

\begin{figure}[H]
\includegraphics[height=2in]{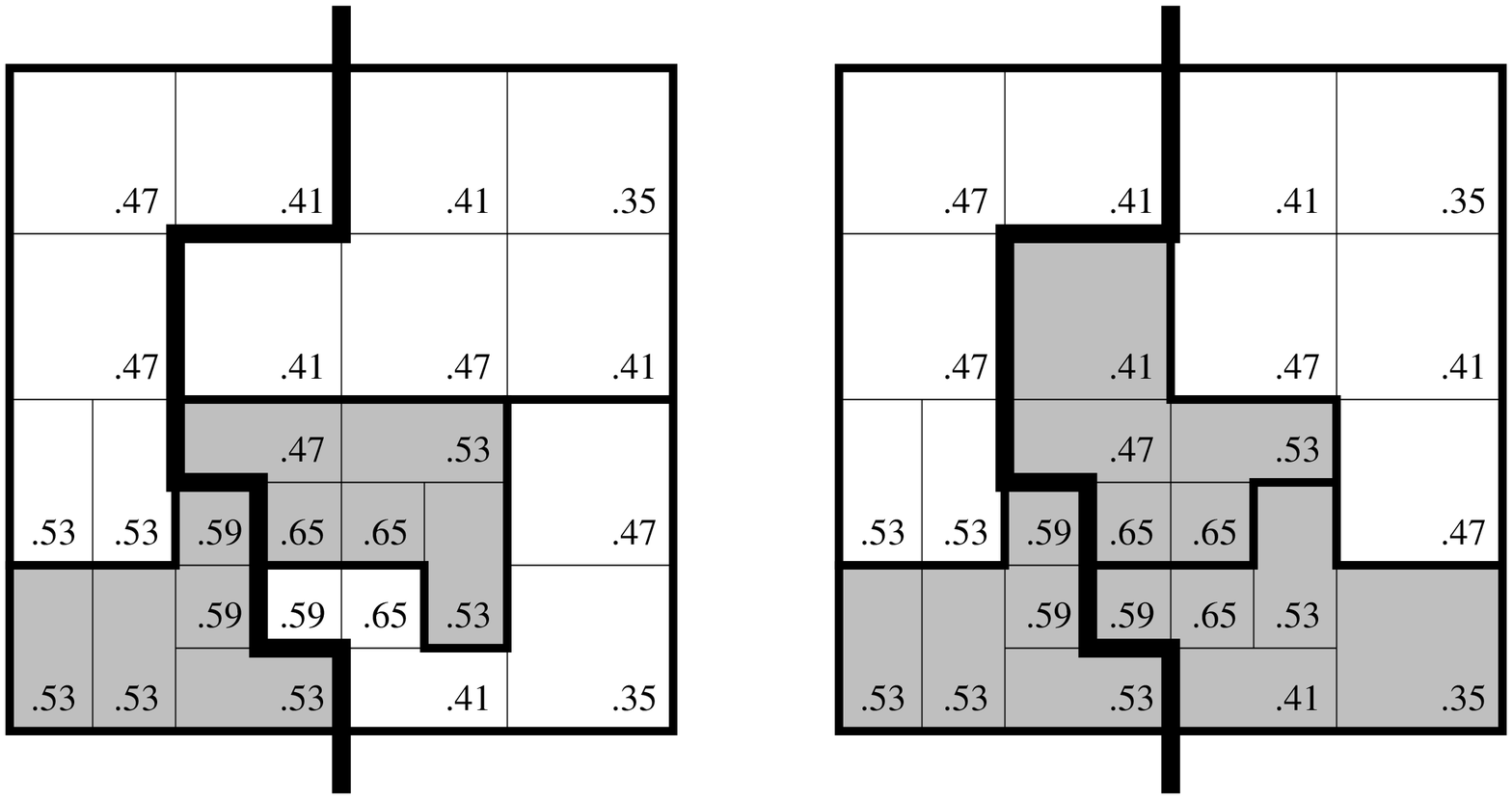}
\caption{Two options for the $2$-split.  The first diagram is option (\ref{optiona}): 
in which $A$ divides the left piece and $B$ divides the right piece.  
The second diagram is option (\ref{optionb}) in which $B$ divides the left piece and $A$ divides the right piece.}
\label{2-split}
\end{figure}

We then consider the same question for the 2-split.  Figure \ref{2-split} shows one possible way each party can optimize its own interests in each of the two options.
Here, party $A$ will still prefer  option (\ref{optionb})  and party $B$ would still prefer option (\ref{optiona}).

\begin{figure}[H] 
\includegraphics[height=2in]{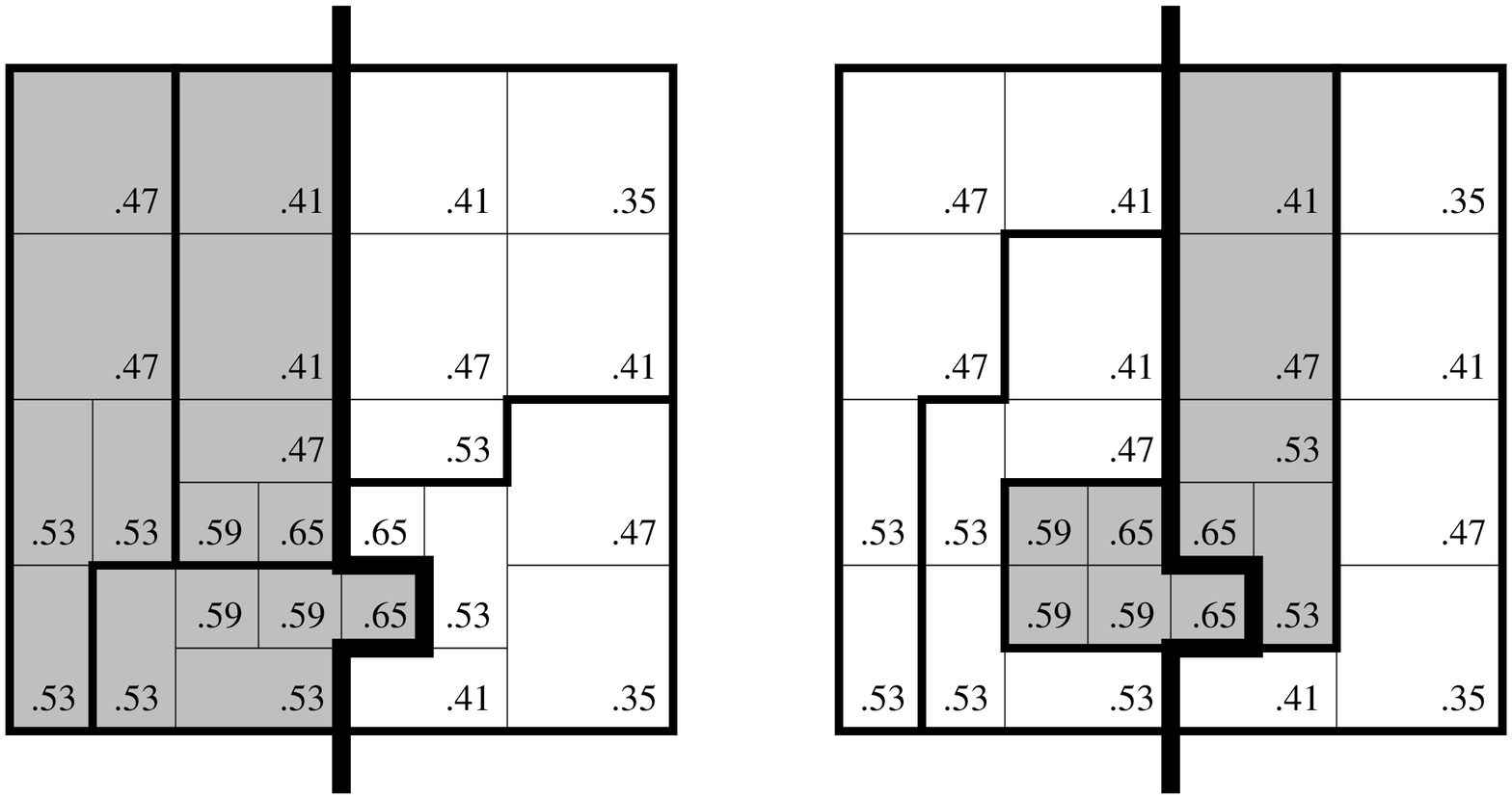}
\caption{Two options for the $3$-split.  The first diagram is option (\ref{optiona}): 
in which $A$ divides the left piece and $B$ divides the right piece.  
The second diagram is option (\ref{optionb}) in which $B$ divides the left piece and $A$ divides the right piece.}
\label{3-split}
\end{figure}

For the 3-split, Figure \ref{3-split}  shows one possible way each party can optimize its own interests in each of the two options.
Notice that now the parties preferences have changed:  party $A$ now prefers (\ref{optiona}) and party $B$ prefers
(\ref{optionb}).

\begin{figure}[H] 
\includegraphics[height=2in]{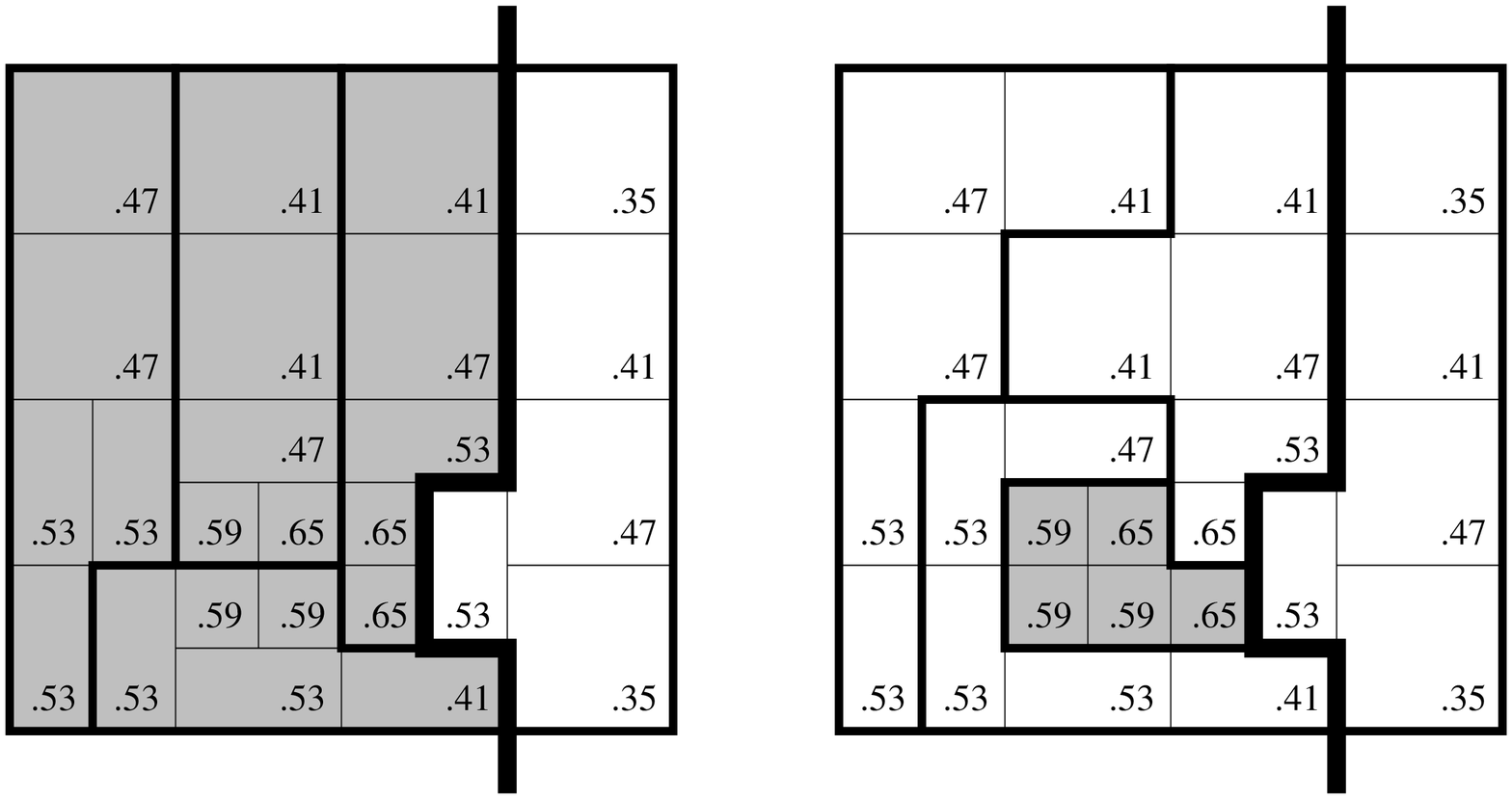}
\caption{Two options for the $4$-split.  The first diagram is option (\ref{optiona}): 
in which $A$ divides the left piece and $B$ divides the right piece.  
The second diagram is option (\ref{optionb}) in which $B$ divides the left piece and $A$ divides the right piece.}
\label{4-split}
\end{figure}

Finally, for the 4-split, Figure \ref{4-split} shows one possible way each party can optimize its own interests in each of the two options.  Again, party $A$ prefers (\ref{optiona}) and party $B$ prefers (\ref{optionb}).  

We summarize the results from party $A$'s point of view in the following table:

\begin{table}[H]
\caption{The number of districts $A$ wins on the left/right sides of split and  total.}
\label{districtswon}

\bigskip
\begin{tabular}{|c | p{2in} | p{2in} |}
\hline
& 
option (\ref{optiona}): left/right=total 
& 
option (\ref{optionb}): left/right = total \\
\hline
1-split &  \hfill 1/1=2 \hfill \ & \hfill 1/3=4 \hfill \  \\
\hline
2-split &  \hfill 1/1=2 \hfill \ & \hfill 1/2=3 \hfill \  \\
\hline
3-split &  \hfill 3/0=3 \hfill \ & \hfill 1/1=2 \hfill \  \\
\hline
4-split & \hfill 4/0=4 \hfill \ & \hfill 1/0=1 \hfill \  \\
\hline
\end{tabular}
\bigskip
\end{table}

With all the preferences now stated, we are ready to move to the final step of the protocol.

{\it Resolution Step.}  Since the two parties prefer different options in each of the four splits, we find the point at which party A's preference switches; here it occurs between the 2-split and the 3-split.  Thus $i_0=2$ and we randomly choose between the four prescriptions corresponding to the options listed in the second and third row of Table \ref{districtswon}.

\begin{enumerate}
\item[i.]  option (\ref{optiona}) for the $2$-split with the result: party $A$ wins 2 districts, party $B$ wins 3.
\item[ii.]option (\ref{optionb}) for the $2$-split with the result: party $A$ wins 3 districts, party $B$ wins 2.
\item[iii.] option (\ref{optiona}) for the $3$-split with the result: party $A$ wins 3 districts, party $B$ wins 2.
\item[iv.] option (\ref{optionb}) for the $3$-split with the result: party $A$ wins 2 districts, party $B$ wins 3.
\end{enumerate}

Notice that these results have party $A$ winning either $40 \%$ or $60 \%$ of the districts, 
the two closest achievable percentages to both the percentage of votes for party $A$ ($50.12\%$) and the percentage of districts given by the absolute geometric target ($2.5/5 = 50 \%$).
This is the result that the protocol is designed to produce; it is argued in  \cite{landau}   that a rigorous result establishing a {\it good choice property} of the protocol combined with the way $i_0$ is chosen will result in this kind of behavior for most choice of split sequences.  We discuss this in the next section.


{\it Augmentation Step.}   For our example, we run the same protocol for the following four additional split sequences:

\begin{figure}[H] 
\includegraphics[height=1.25in]{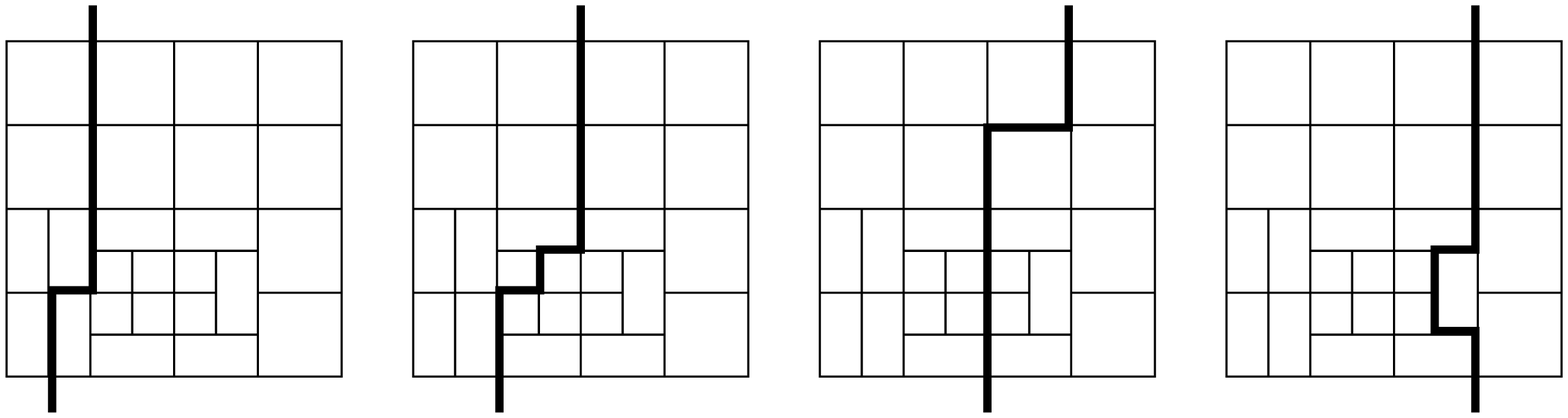}
\caption{Vertical Split Sequence.}
\end{figure}

\begin{figure}[H]  
\includegraphics[height=1.01in]{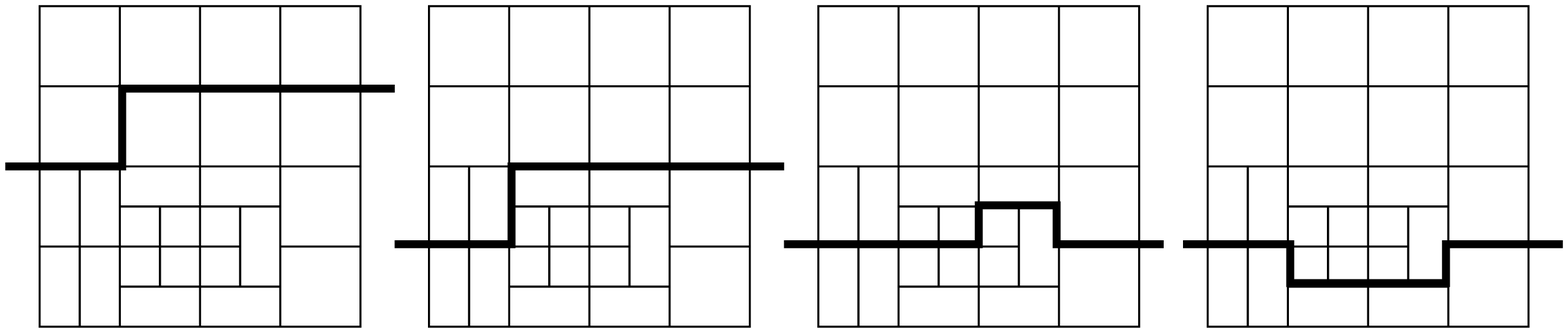}
\caption{Horizontal Split Sequence.}
\end{figure}

\begin{figure}[H] 
\includegraphics[height=1.25in]{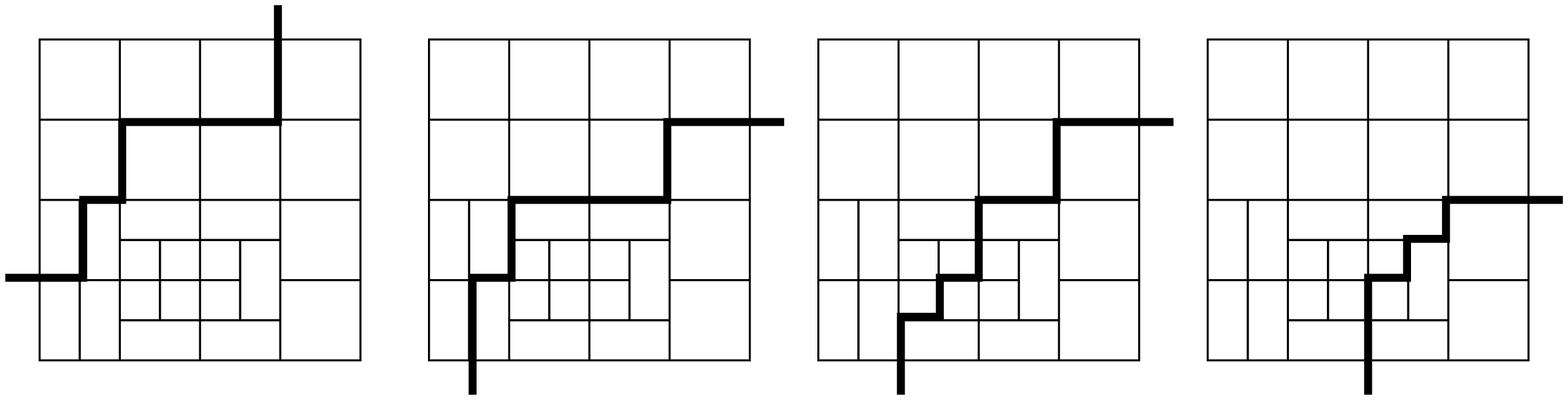}
\caption{Diagonal Split Sequence 1.}
\end{figure}

\begin{figure}[H] 
\includegraphics[height=1.25in]{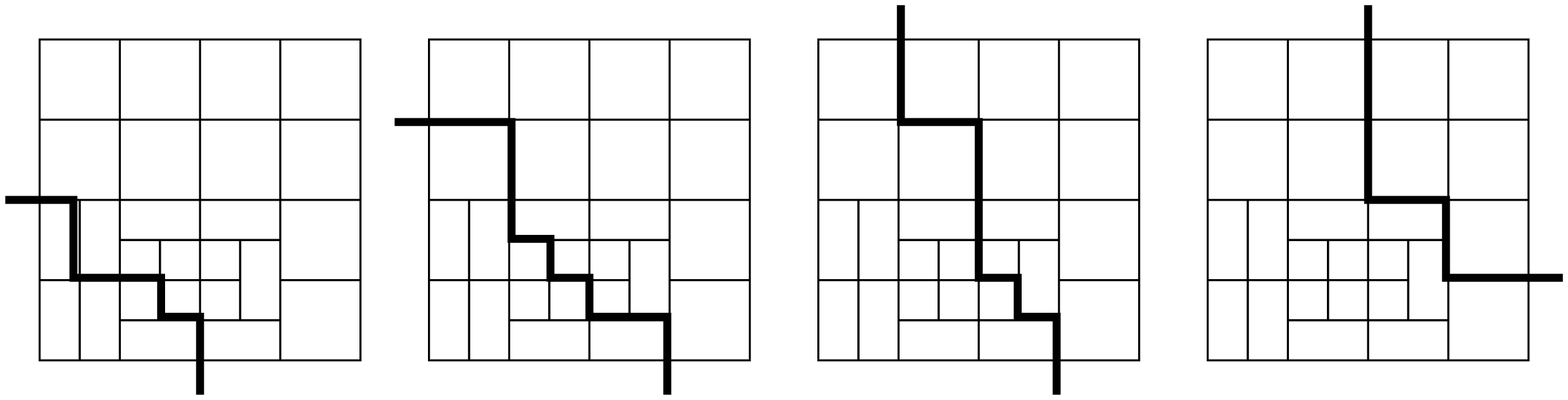}
\caption{Diagonal Split Sequence 2.}
\end{figure}


These have the following outcomes from party $A$'s perspective, listed in Table \ref{bigtable}.

\begin{table}[H]
\caption{The number of districts $A$ wins in piece 1, piece 2, and total.}
\label{bigtable}
\medskip

{\small
\begin{tabular}{|c|c | p{2in} | p{2in} |}
\hline
 
&& 
option (\ref{optiona}): pc.1/pc.2 = total & 
option (\ref{optionb}): pc.1/pc.2 = total \\
\hline
\multirow{4}{*}{Vertical Split} 
&1-split &  \hfill 1/1=2 \hfill \ & \hfill 1/3=4 \hfill \  \\
\cline{2-4}
 &2-split &   \hfill 1/1=2 \hfill \ & \hfill 1/2=3 \hfill \  \\
\cline{2-4}
&3-split &  \hfill 2/0=2 \hfill \ & \hfill 1/1=2 \hfill \  \\ 
\cline{2-4}
&4-split & \hfill 4/0=4 \hfill \ & \hfill 1/0=1 \hfill \  \\
\hline
\multirow{4}{*}{Horizontal Split} 
&1-split &  \hfill 0/1=1 \hfill \ & \hfill 0/4=4 \hfill \  \\
\cline{2-4}
 &2-split &   \hfill 0/2=2 \hfill \ & \hfill 0/3=3 \hfill \  \\
\cline{2-4}
&3-split &  \hfill 2/1=3 \hfill \ & \hfill 1/2=3 \hfill \  \\ 
\cline{2-4}
&4-split & \hfill 3/0=3 \hfill \ & \hfill 1/0=1 \hfill \  \\
\hline
\multirow{4}{*}{Diagonal Split 1} 
&1-split &  \hfill 0/1=1 \hfill \ & \hfill 0/3=3 \hfill \  \\
\cline{2-4}
 &2-split &   \hfill 1/1=2 \hfill \ & \hfill 0/3=3 \hfill \  \\
\cline{2-4}
&3-split &  \hfill 2/1=3 \hfill \ & \hfill 1/2=3 \hfill \  \\ 
\cline{2-4}
&4-split & \hfill 3/0=3 \hfill \ & \hfill 1/0=1 \hfill \  \\
\hline
\multirow{4}{*}{Diagonal Split 2} 
&1-split &  \hfill 1/1=2 \hfill \ & \hfill 1/3=4 \hfill \  \\
\cline{2-4}
 &2-split &   \hfill 2/0=2 \hfill \ & \hfill 2/2=4 \hfill \  \\
\cline{2-4}
&3-split &  \hfill 3/0=3 \hfill \ & \hfill 2/1=3 \hfill \  \\ 
\cline{2-4}
&4-split & \hfill 4/0=4 \hfill \ & \hfill 1/0=1 \hfill \  \\
\hline
\end{tabular}
}
\end{table}

Unlike the first split sequence from Figure \ref{newsplit} that  we explored in detail, the result of each of these split sequences is that the parties will be indifferent to one split.  Here are possible maps from the results of the protocol for each of these split sequences:

\begin{figure}[H] 
\includegraphics[height=2in]{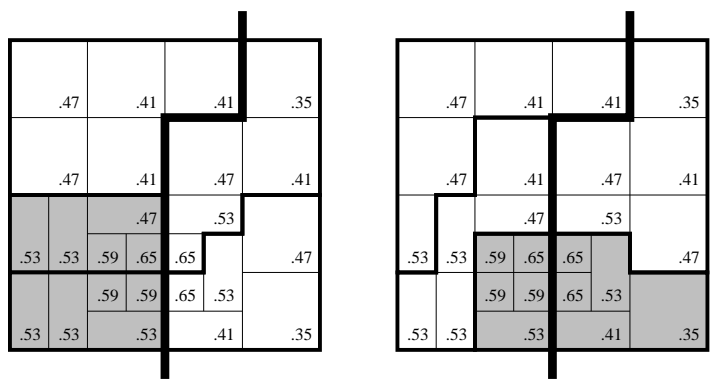}
\caption{Outcomes for Vertical Split Sequence.}
\end{figure}

\begin{figure}[H]  
\includegraphics[height=1.7in]{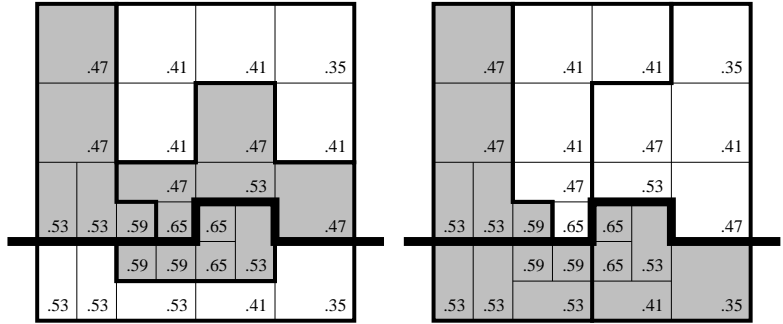}
\caption{Outcomes for Horizontal Split Sequence.}
\end{figure}

\begin{figure}[H] 
\includegraphics[height=1.84in]{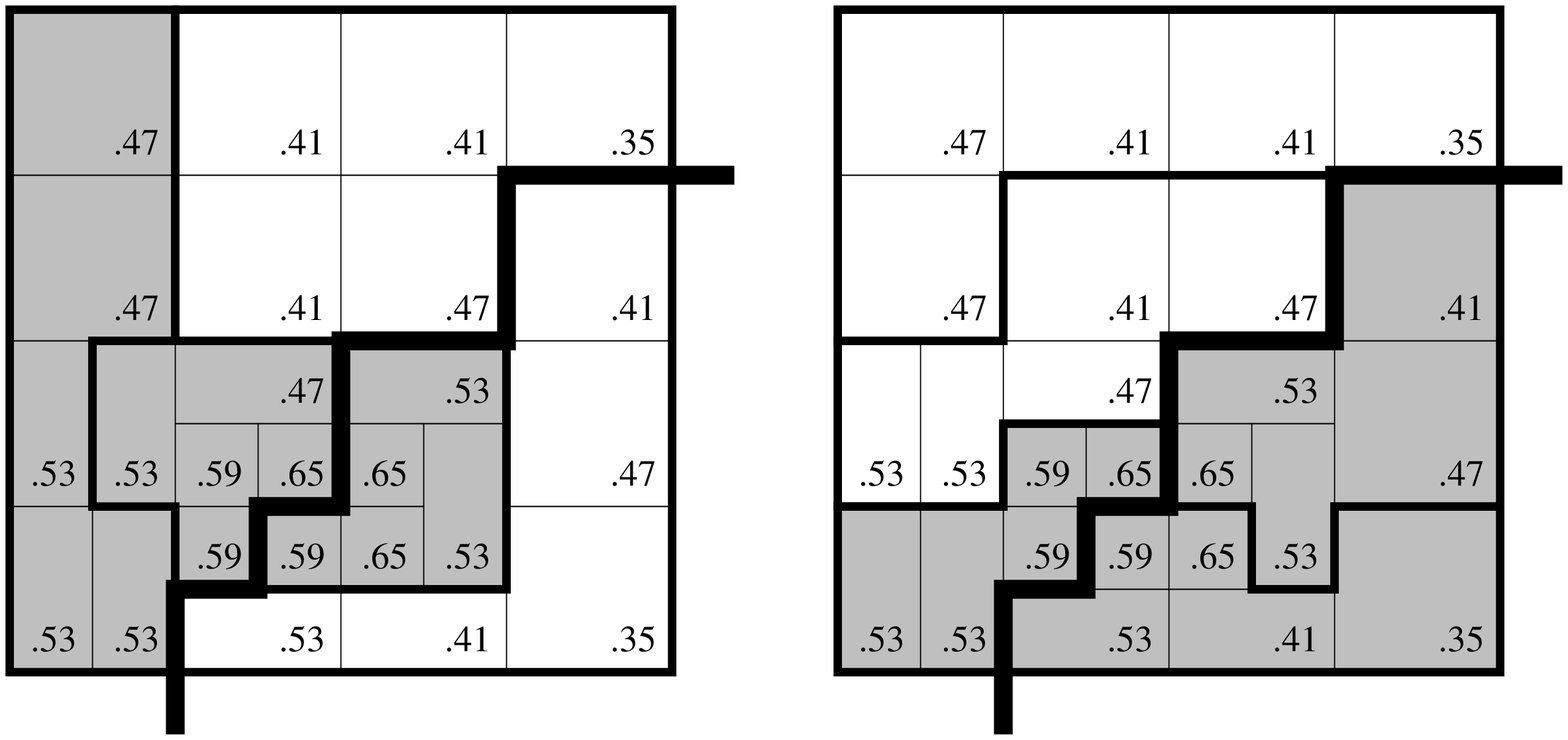}
\caption{Outcomes for Diagonal Split Sequence 1.}
\end{figure}

\begin{figure}[H] 
\includegraphics[height=2in]{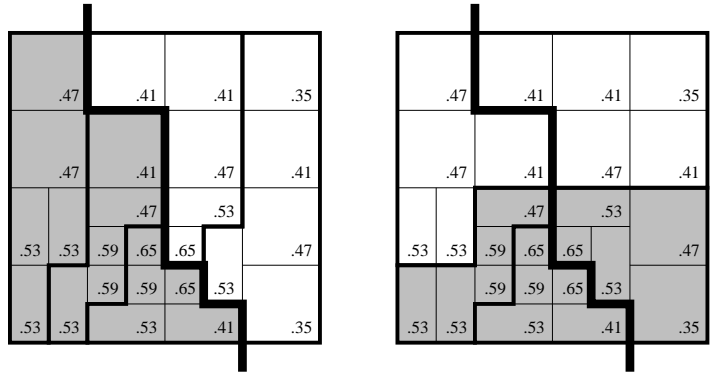}
\caption{Outcomes for Diagonal Split Sequence 2.}
\end{figure}


In the ranking protocol, both parties would rank the 5 different outcomes from best to worst.  Depending on the outcome of the random choice of prescription for the first split sequence (Figure \ref{newsplit}), the outcomes would be party $A$ winning 3 districts 3 or 4 times (Horizontal Split sequence, both Diagonal split sequences, and possibly the first split sequence),  and winning 2 districts 1 or 2 times (Vertical split sequence, and possibly the first split sequence).  Since the result of the ranking protocol will result in one of the top three outcomes for both parties, in this case, the final outcome will be party $A$ winning 3 districts and party $B$ winning 2.  We remark that this resolution is as close as one can get to both the absolute geometric target (2.5) and the proportion of constituent voters (50.12 \%)

\subsection{Fairness qualities of the protocol}

Having explained the protocol we turn to a discussion of why the protocol is fair.   We will analyze the protocol from the point of view of party $A$; the identical analysis can be made for party $B$.  We address the following two questions:

\begin{itemize}
\item {\it  If the map is created from a choice party A preferred or was indifferent to, will it be fair for party A?}  


\item {\it What if the randomness in the algorithm results in a choice party A did not prefer?}
\end{itemize}

The first question is answered in the affirmative by establishing that the protocol has the  {\it good choice property} \cite{landau}.  Approximately,  the good choice property says that if party $A$ is using a voting model $V$ and has an additive rating system $R$, there will be a choice for party $A$ that achieves an outcome that is at least as good as a number close to the geometric target for $V$ and $R$ (see Section \ref{s:fairshare} for definitions).   Precisely, for a given $k$-split define {\it the party's  $k$-split geometric target for $V$ and $R$} to be the average rating of the best and worst outcomes over all viable divisions {\it that include} the dividing line of the $k$-split.  Then the good choice property is:

\begin{thm}[Good Choice Property \cite{landau}]
For any voting model $V$ and rating system $R$,  one of the choices given in the protocol achieves an outcome that is at least as good as the party's $k$-split geometric target for $V$ and $R$.  
\end{thm}

\begin{figure}[h] 
\includegraphics[height=3in]{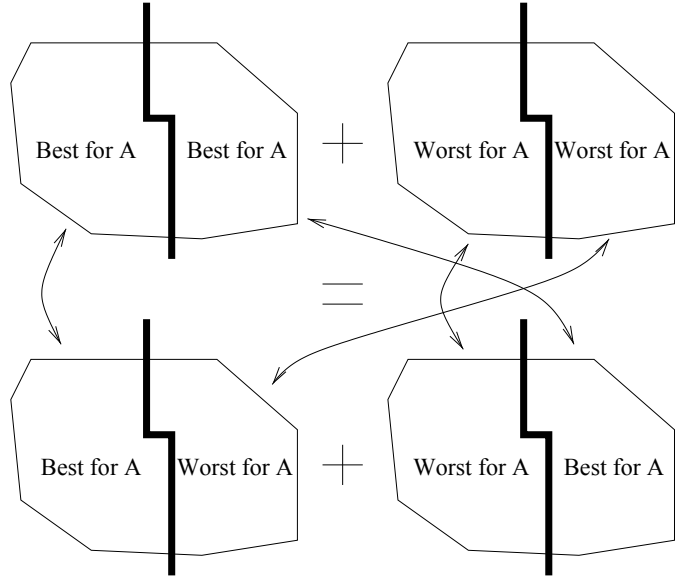}
\caption{The good choice property.}
\label{goodchoice}
\end{figure}

For our example above party A has been acting according to its interests with $V=V_{out}$ and $R=R_{win}$. The good choice property follows from the following observation that is perhaps best seen pictorially (see Figure \ref{goodchoice}):  given a particular $k$-split, the average of the number of districts won by party $A$ under options (\ref{optiona}) and (\ref{optionb}) is equal to the average of the number of districts that party $A$ wins if it had complete control (which would result in the best outcome for party $A$) and if it had no control (which would result in an outcome no worse than the worst for party $A$).  

Thus at least one of the two options is better than the average outcome between the best and worst scenario for party $A$, which is precisely the definition of the $k$-split geometric target when $V=V_{out}$ and $R=R_{win}$.   It should be clear that the same argument holds regardless of choice of $V$ and additive rating system $R$.

The astute reader will  notice that the $k$-split geometric target for $V$ and $R$ can differ from the geometric target for $V$ and $R$; in other words, insisting that the division includes the boundary given by the $k$-split can penalize one party.  Two observation suggest that should this happen, the penalty will not be large.  First, the choice of split is made by an independent (neutral) third party and thus should be no more biased against one party than a random choice.   Second, in the case where there are no geometric constraints (see \cite{landau}), the absolute geometric target and the $k$-split geometric target for $V_{out}$ and $R_{win}$ can differ by at most $\frac{1}{2}$.  In our example, we see this difference between the $k$-split geometric target and the absolute geometric target:  for party $A$, in the 3-split in Vertical Split sequence, and for party $B$, in  the 3-splits on the final three split sequences of the protocol.  In each of these cases, however,  the difference from the absolute geometric target is as small as it could be: $\frac{1}{2}$.   It is reasonable to assume that most splits will either not particularly favor either party, or favor a party by a small amount. It is then the augmentation step of the protocol that ensures that a rare  ``bad'' split for a particular party will not come into play (since the affected party would put such a split towards the bottom of their rankings).

We see therefore, that the good choice property, when coupled with the augmented protocol, implies that party $A$ should be satisfied if the division is created by an option that it chose. 

We now turn to the second question---how party $A$ will fare if the randomness in the algorithm results in a choice they did not prefer. 
The randomness is implemented only if for each $i$-split, the two parties have opposite preferences (for instance in the first split sequence described in Figure \ref{newsplit}).  We suppose the random prescription in the Resolution Step (see Section \ref{s:protocol}) is one not preferred by party $A$, for instance prescription (i.) in the Resolution step (i.e. option (\ref{optiona}) for $i=i_0$).  In our example, this would correspond to $i_0=2$ in the first split sequence.   Party $A$, however, prefered option (\ref{optionb}): to divide piece 2 and have party $B$ divide piece 1.  Notice, however, that party $A$ would prefer to divide up piece 1 in the $i_0+1$ split, and this piece 1 only differs from the piece 1 of the $i_0$ split by a small region with a population equal to the size of a single district.  (Similarly piece 2 in the $i_0$ and $i_0 +1$ splits  only differ by this same small region).  Because party $A$ prefers option (\ref{optionb}) for the $i_0$ split and option (\ref{optiona}) for the $i_0+1$ split (and because piece 1 of these two splits do not differ by very much), it is reasonable to expect that party $A$'s preference for option (\ref{optionb}) over option (\ref{optiona}) for the $i_0$ split is mild.  (In our example, this is indeed the case as party $A$ achieves an outcome of winning 2 districts which is of minimal negative deviation from the absolute geometric target of $2.5$). If indeed this is the case, then party $A$'s discontent with the division would only be mild as we have shown (by the good choice property) that party $A$ would have been satisfied with the slightly better option of (\ref{optionb}). 

Even though the first pieces of the two splits differ by a small amount, one can construct scenarios where that small amount makes a big difference.  However, recall that the creation of the splits was done by an independent party and therefore one would expect this type of scenario to be rare.  Again, choosing to use the augmented protocol would ensure that this rare scenario would not occur in the division chosen.

\section{Conclusion}

Replacing current redistricting procedures with the protocol presented here surely presents substantial political obstacles.  It has been observed numerous times (e.g., see \cite{hirsch-mann-nyt}) that any proposed change should be structured to take effect far enough in the future so that it could not be interpreted as a power grab by one party.  However, as noted in the last paragraph of the introduction, some of the ideas presented here could be incorporated into current processes without requiring a complete overhaul of the redistricting process.

In this article, we have used a detailed example to explore the redistricting protocol of \cite{landau}.
We have shown how this procedure retains the usual constraints that may be desirable to impose on a redistricting solution, while incorporating some of the best features of a fair division procedure: multilateral evaluation, procedural fairness, and fairness guarantees.  Procedural fairness is apparent in the protocol, the geometric targets incorporate multilateral evaluation, and the ability to ensure outcomes near geometric targets provides the fairness guarantee.
The result is a solution that accounts for both parties having different interests, involves a resolution process and an interactive protocol to elicit preferences, and provides mathematical confidence that the outcome will be fair.


\begin{thebibliography}{AA}


\bibitem{app}  N. Apollonio, R. I. Becker, I. Lari, F. Ricca, and B. Simeone, The Sunfish against the Octopus: opposing compactness to gerrymandering, in \emph{Mathematics and Democracy.  Recent advances in Voting Systems and Collective Choice, Studies in Choice and Welfare} (2006), 19--41.

\bibitem{brams-jones-klamler} S. J. Brams, M. A. Jones and C. Klamler, $N$-Person Cake-Cutting: There May Be No Perfect Division, \emph{Amer. Math. Monthly} {\bf 120} (2013), 35--47.

\bibitem{brams-taylor-book} S. J. Brams and A. D. Taylor, {\em Fair
Division: from Cake-Cutting to Dispute Resolution},  Cambridge
University Press, 1996 .

\bibitem{brams-taylor-win} S. J. Brams and A. D. Taylor, {\em The Win-Win Solution: Guaranteeing Fair Shares to Everybody},  New York: 
W. W. Norton, Inc., 1999.



\bibitem{hirsch-mann-nyt} S. Hirsch and T. E. Mann, ``For Election Reform, a Heartening Defeat'', \emph{New York Times}, Nov.~11, 2005, also at {\tt http://www.nytimes.com/2005/11/11/opinion/11mann.html}.

\bibitem{landau} Z. Landau, O. Reid and I. Yershov,  A Fair Division Solution to the Problem of Redistricting, {\em Social Choice and Welfare} {\bf 32}, Issue 3 (2009), 479--492.

\bibitem{moulin}
H. Moulin, {\em Fair Division and Collective Welfare}, MIT Press, 2003.

\bibitem{electstat} Office of the Clerk, U.S. House of Representatives website: \\
{\tt http://clerk.house.gov/members/electionInfo/elections.html}

\bibitem{peterson-su} E. Peterson and F. E. Su, Four-person envy-free chore division,
\emph{Math. Mag.} {\bf 75} (2002), 117---122. 


\bibitem{rawls} J. Rawls, \emph{A Theory of Justice}, Harvard University Press, 1971.

\bibitem{redlaw2000} Redistricting Law 2000. Denver, CO: National Conference of State Legislatures, 1999. Available at \\{\tt http://www.senate.leg.state.mn.us/departments/scr/redist/} \\ {\tt red2000/red-tc.htm}

\bibitem{robertson-webb} J. M. Robertson and W. A. Webb, {\em Cake-Cutting 
Algorithms: Be Fair If You Can}, A. K. Peters Ltd., 1998.



\bibitem{spector}
B. I. Spector, Analytical Support to Negotiations: An Empirical Assessment,
\emph{Group Decision and Negotiation} {\bf 6} (1997), 421--436.

\bibitem{steinhaus} H. Steinhaus, The problem of fair division,
\emph{Econometrica} {\bf 16} (1948), 101--104.

\bibitem{su} F. E. Su, Rental harmony: Sperner's lemma in fair division,
\emph{Amer. Math. Monthly} {\bf 106} (1999), 930--942. 



\end{thebibliography}
\end{document}